\documentclass[twoside]{article}

\usepackage{url}
\usepackage{wrapfig,rotating}
\usepackage{amssymb,amsmath,array}
\usepackage{graphics,epsfig,color}
\usepackage{graphicx}

\begin{document}

\hfill MZ-TH/09-31 

\parindent 0mm
\null
\vskip 2em
{\bfseries\LARGE Photon 2009: Summary of Theory Talks\footnote{%
  {\em Photon09}: International Conference on the Structure and the 
  Interactions of the Photon including the 18th International Workshop 
  on Photon-Photon Collisions and the International Workshop on High 
  Energy Photon Linear Colliders, 11-15 May 2009, DESY, Hamburg}
 \par}
\vskip 1.5em
{\lineskip .5em
   {\slshape Hubert Spiesberger}\\[1ex]
   Institut f\"ur Physik - Johannes-Gutenberg-Universit\"at \\
   Staudinger Weg 7, D-55099 Mainz, Germany
}

\begin{abstract}
The conference {\em Photon 2009} on the structure and the interactions 
of the photon included sessions on photon-photon collisions and a 
future high-energy photon linear collider. This summary of theoretical 
contributions to the conference therefore has two parts. I will 
discuss the physics potential of photon colliders with an emphasis on 
the study of electroweak physics and the search for physics beyond 
the standard model. Secondly, I will describe a few highlights in 
recent progress in the understanding of the properties and the 
interaction of the photon, comprising the production of prompt 
photons, the photon structure and exclusive hadron production, 
small-$x$ and total cross sections of deep inelastic scattering. 
Finally, I will review the status of the comparison of measurement 
and theory for the muon anomalous magnetic moment $g-2$. 
\end{abstract}

\section{Introduction}

The photon, the leitmotif of the conference {\em Photon 2009}, has 
appeared in the various talks in two different roles. First, the 
photon is an ideal tool to probe new physics in high-energy 
photon-photon or photon-electron collisions. During recent years a 
lot of work has been devoted to the study of the prospects of photon 
colliders in the search for physics beyond the standard model or to 
perform precision measurements of standard model phenomena. It was 
also suggested that a photon collider would be the suitable place 
to study the properties of a Higgs boson, if it had a mass of about 
120 GeV. In particular, the possibility to build a photon-collider 
as a precursor of an electron-positron linear collider has been 
discussed, but was not supported by the International Linear Collider 
Steering Committee \cite{telnov} and the construction of such a 
facility will have to wait for its turn, maybe as an extension of an 
$e^+e^-$ linear collider. Corresponding plans will have a chance to 
be revived only, if results of forthcoming experiments at the LHC 
would require new theoretical ideas whose further scrutiny at a photon 
collider can be expected to considerably improve our knowledge. Till 
then, it may be worthwhile to study the feasibility of experimentation 
with $\gamma \gamma$ and $\gamma p$ collisions at the LHC.

Secondly, as the main and traditional subject of this series of 
conferences, the photon has appeared as a research object on its 
own right. The study of the photon and its properties, described in 
terms of structure functions that are measured first of all in 
virtual-photon scattering (i.e., in deep-inelastic lepton-nucleon 
scattering), and its generalizations needed for a description of 
exclusive processes, opens a wide field of research topics in QCD, 
located at the border-line of perturbative and non-perturbative 
phenomena. There is a wealth of experimental data not yet understood 
at a quantitative level and more theoretical work is needed. The 
photon as a theory laboratory, as an object to gain a deeper 
understanding of theoretical concepts, as for example factorization 
in exclusive processes, is often helpful. Recent progress in this 
field of research is highlighted in a second part of this summary. 

A topic of special importance where aspects of both electroweak and 
strong interactions play a crucial role, is the measurement of the 
anomalous magnetic moment of the muon and the comparison with 
theoretical predictions. A short status review is presented at the 
end of this article. 

This printed version of the summary talk includes only a tiny portion 
of the figures that have been shown at the conference. The text should 
therefore be read together with the slides \cite{tsummary}.

\section{Physics potential of a photon collider}

\subsection{Motivation}

\begin{wrapfigure}{r}{0.63\columnwidth}
\centerline{\includegraphics[width=0.6\columnwidth]
{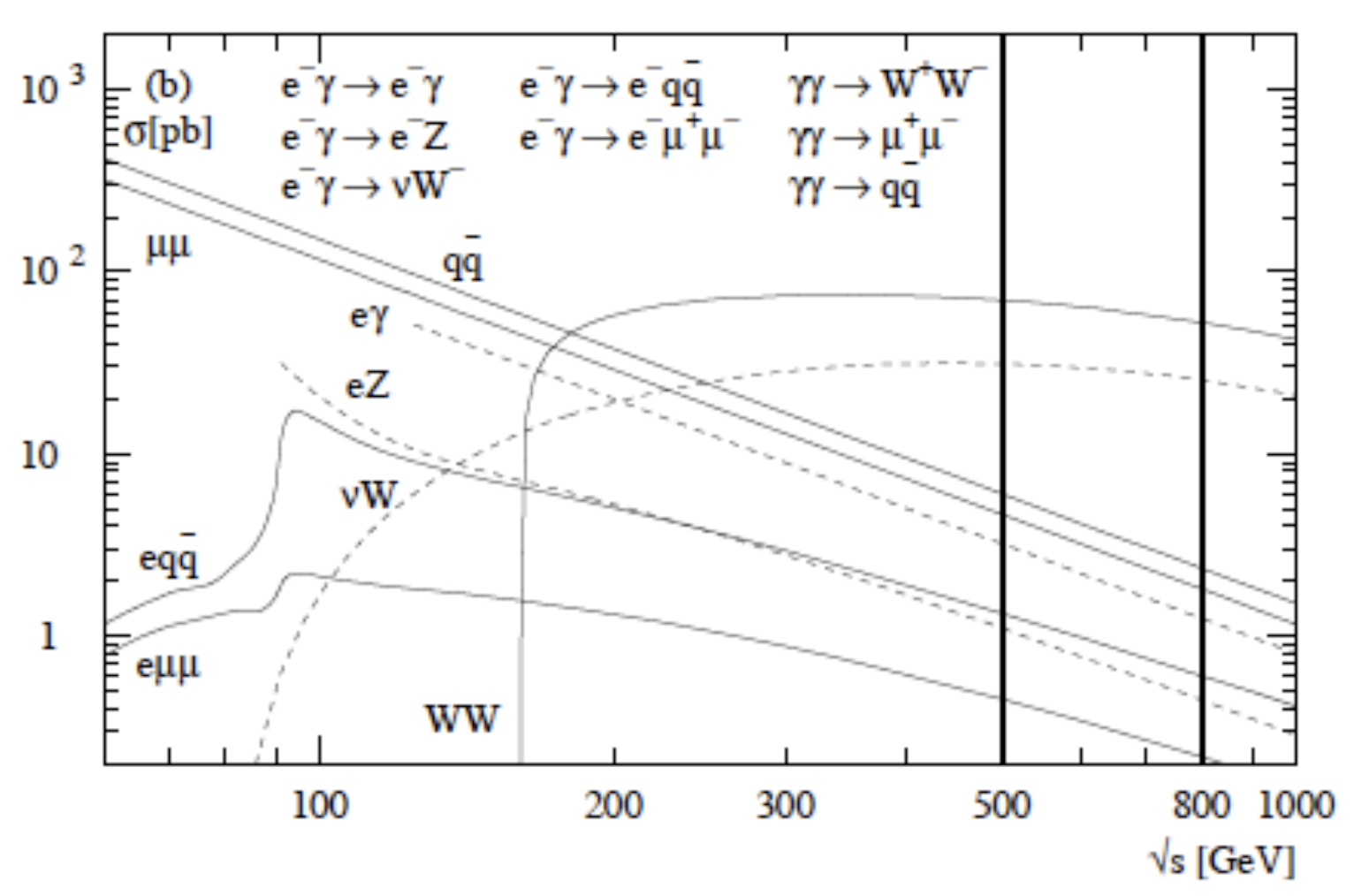}}
\caption{Typical cross sections at $e\gamma$ and $\gamma \gamma$ 
colliders \cite{ggxsec}.}
\label{Fig:hs-1}
\end{wrapfigure}

A first motivation for the study of scattering processes with one or 
two photons in the initial state comes from a naive evaluation of 
typical cross sections for standard model processes. These cross 
sections are often comparable, in some cases even larger, than for 
corresponding processes in $e^+ e^-$ annihilation. From a quick look 
at Fig.\ \ref{Fig:hs-1} it should be obvious that detailed studies 
including experimental conditions are worthwhile being performed. In 
particular one may expect that through studies of the $W^+ W^-$ final 
state one can obtain information on 3- and 4-boson interactions. A 
crucial question is, however, whether photon collisions can be 
realized with high enough luminosity and present technical design 
studies indicate that this would indeed be the main limitation. It is 
therefore important to identify possible measurements at a photon 
collider that provide information complementary to what can be found 
at an $e^+ e^-$ linear collider. It was argued that in particular with 
respect to the determination of properties of standard model or 
supersymmetric Higgs bosons a photon collider may be advantageous in 
comparison with $e^+e^-$ collisions\footnote{For more details 
  and for phenomenological aspects of QCD and hadron physics 
  at photon colliders, see Refs.\ \cite{ggxsec,tesla}.}. 

For a meaningful assessment of the possible reach of high energy 
experiments in the seach for new phenomena, it is an important 
prerequisite to know the constraints on model parameters imposed 
by present day's experiments, including those at low energies. A new 
tool has been presented \cite{mahmoudi} that allows to obtain allowed 
parameter ranges for supersymmetric models and general 2-Higgs-doublet 
models. Flavor physics observables like rare $B$ decays are viewed 
as most promising for this purpose.  

\subsection{Higgs bosons at a photon collider}

It is a common believe that a Higgs boson, if it exists, can not 
be missed at the forthcoming experiments at the LHC. The question 
whether an observed Higgs boson fits into a supersymmetric model 
will then be of utmost importance. In the minimal supersymmetric 
standard model (MSSM), there is a large region in parameter space
where the LHC will not be able to distinguish a SM from a MSSM 
Higgs boson. This 'blind wedge' covers values above $m_A \simeq 200$ 
GeV, the mass of the $CP$-odd Higgs particle, and a region around 
$\tan \beta \simeq 5$ increasing in size with increasing $m_A$. 
Precise measurements of the decay branching ratios will then be 
needed. In particular if $CP$-violating interactions are present in 
the Higgs sector, a potential photon collider will provide additional 
help to identify the correct underlying theory. 

\begin{wrapfigure}{r}{0.53\columnwidth}
\centerline{\includegraphics[width=0.4\columnwidth]
{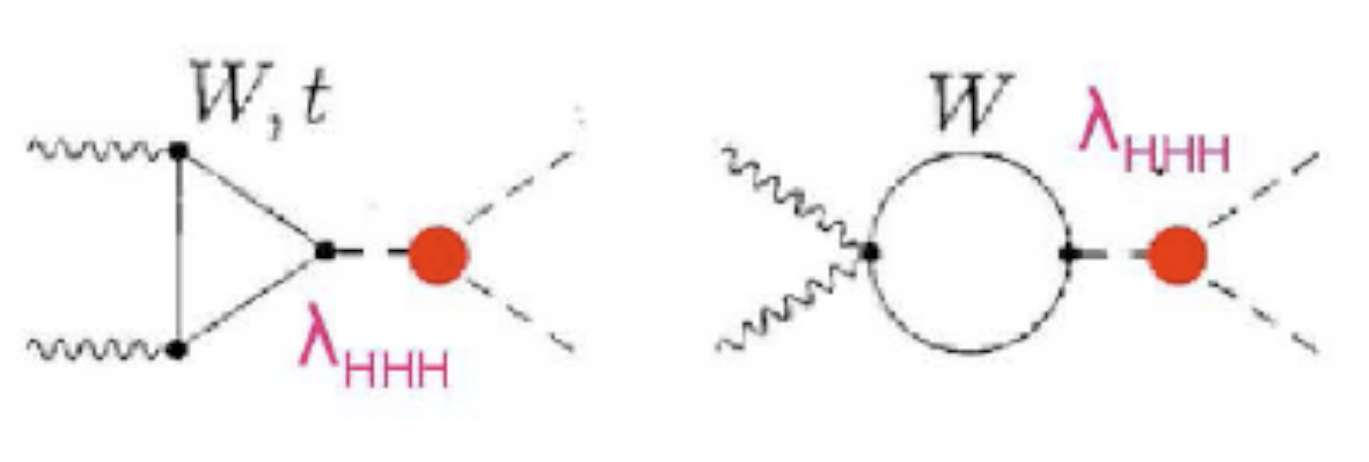}}
\centerline{\includegraphics[width=0.52\columnwidth]
{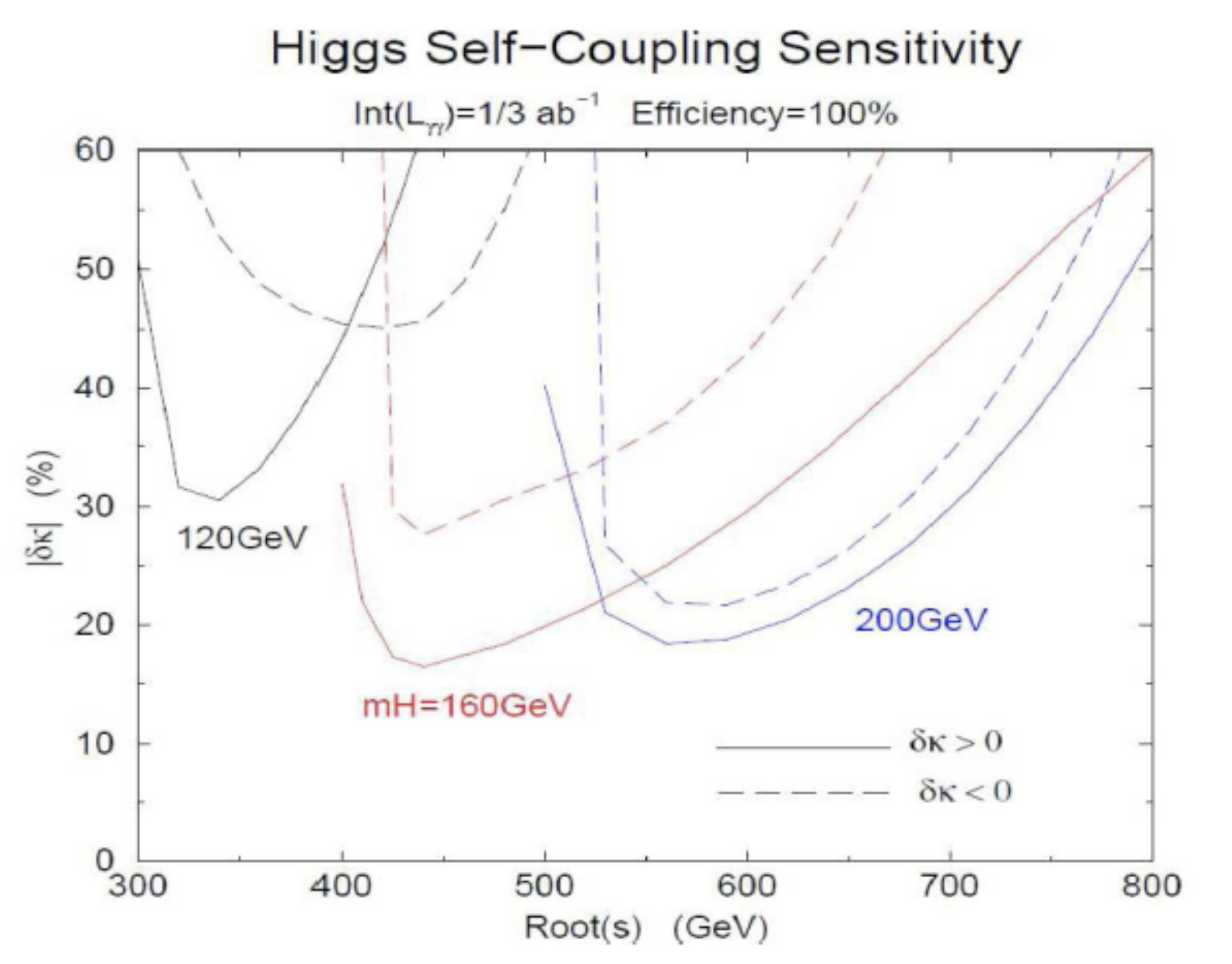}}
\caption{Higgs boson self coupling: sensitivity at a $\gamma \gamma$ 
collider \cite{h3sens}.}
\label{Fig:hs-2}
\end{wrapfigure}

In $\gamma \gamma$ collisions, $s$-channel production of the 
electrically neutral SM Higgs boson proceeds via a triangle loop of 
the heaviest SM particles, mainly top and $W$. A measurement of the 
production cross section will provide a determination of the 2-photon 
decay width $\Gamma(H \rightarrow \gamma \gamma)$ \cite{spira}. 
Realistic studies have shown that a statistical precision of roughly 
2\,\% can be reached for a light Higgs boson in the interesting mass 
range of $M_H = 120$ GeV with a total luminosity of 410 fb$^{-1}$ in 
the decay channel $H \rightarrow b \bar{b}$. For a heavy $CP$-even or 
$CP$-odd MSSM Higgs boson, the precision is worse, ranging between 10 
and 20\,\% \cite{niezurawski}. If the energy of $\gamma \gamma$ 
collisions is high enough, the production of two Higgs bosons will 
be observable and cross section measurements will allow one to 
determine also the values of the Higgs boson self couplings. 
Corresponding information will be crucial to an analysis of the Higgs 
sector. At an $e^+ e^-$ linear collider, a precision between 10 and 
20\,\% for a measurement of the three-Higgs coupling $\lambda_{HHH}$ 
can be reached, provided the center-of-mass energy is $\sqrt{s} = 1$ 
TeV and $M_H$ in the range between 100 and 200 GeV. At lower $\sqrt{s} 
= 500$ GeV, $\Delta \lambda_{HHH} / \lambda_{HHH}$ will degrade 
quickly, in particular for larger $M_H$. For larger values of the 
Higgs mass up to $M_H \simeq 200$ GeV, measurements of the process 
$\gamma \gamma \rightarrow HH$ are expected to provide a determination 
of $\Delta \lambda_{HHH} / \lambda_{HHH}$ with a statistical precision 
of about 20\,\%, even in the energy range between $\sqrt{s} \simeq 
520 - 650$ GeV, assuming 100\,\% tagging efficiency (see Fig.\ 
\ref{Fig:hs-2}).

In models which implement $CP$ violation in the Higgs sector, a photon 
collider will play a crucial role \cite{godbole}. In the presence of 
$CP$ violation, the three neutral states in the SUSY Higgs sector may 
mix with no fixed $CP$ property. The coupling to the $Z$ boson, 
relevant for the dominating discovery channel $e^+ e^- \rightarrow Z 
\phi_1$ may be reduced for the lightest scalar mass eigenstate 
$\phi_1$. In fact, the search at LEP may have missed such Higgs bosons 
and existing limits, for example from OPAL, can not exclude the full 
range of $\tan \beta$ even for masses in the range below 50 GeV. It 
may also be missed at the Tevatron and at the LHC if the coupling to 
top quarks is also reduced. A photon collider would be the place for 
a discovery since there these states can always be produced. The 
possibility to control the polarization of the back-scattered photons 
will be an important prerequisite for a study of the $CP$ properties 
of Higgs bosons; an analysis of $WW$ and $ZZ$ final states in 
combination with information obtained from top (and $\tau$) 
polarization observables may then provide access to regions in the 
parameter space of $CP$-violating scenarios which are unreachable by 
the LHC. Also mixed polarization-charge asymmetries in $e\gamma$ 
collisions with a polarized electron beam have been studied as a means 
to enhance the signal of heavy $CP$-even or $CP$-odd Higgs production 
\cite{godbole}. 

In the MSSM or in general 2-Higgs-doublet models, violation of $CP$ 
can occur via different mechanisms. As is well-known in the case of 
$CP$ violation in the quark sector of the standard model, a careful 
definition of parameters is needed, even more so in the Higgs sector 
in order to identify the specific mechanism. Reparametrization 
invariants, similar to the Jarlskog determinant, can be defined and 
allow to distinguish $CP$ violation through mixing of states or 
through direct $CP$-violating interactions at tree level 
\cite{krawczyk}.

A characteristic feature of the MSSM is that there is decoupling of 
the heavy states in the Higgs sector, i.e.\ the MSSM becomes 
indistinguishable from the SM if the masses of the heavier Higgs 
bosons (and of the superpartner particles, of course) become large. 
In a general 2-Higgs doublet model this is not necessarily the case. 
Even if the model parameters are chosen such that all tree-level 
couplings to fermions and gauge bosons are the same as in the standard 
model, differences may enter through one-loop contributions. The 
production of a pair of the lightest $CP$-even Higgs boson $h^0$ in 
$\gamma \gamma$ collisions proceeds both in the standard model and in 
its extensions with two Higgs doublets via one-loop diagrams and is 
therefore an ideal place to look for differences. An important 
contribution to this process comes from one-loop corrections to the 
$h^0 h^0 h^0$ vertex which is affected by non-decoupling effects: 
after renormalization, quartic mass terms of the heavier Higgs bosons 
remain and lead to enhancements with respect to the tree-level 
coupling. These non-decoupling effects may be visible in the 
production cross section $\sigma(\gamma \gamma \rightarrow h^0 h^0)$.  
Also diagrams with the charged Higgs boson contribute here and 
depending on its mass $m_{H^{\pm}}$, cross sections can be orders of 
magnitudes above the (MS)SM prediction \cite{rsantos,corholl}. More 
exotic extensions of the Higgs sector may predict the existence of 
doubly charged Higgs bosons. The production process $\gamma\gamma 
\rightarrow H^{++}H^{--}$ is enhanced by the square of the charge as 
compared to production in $e^+e^-$ annihilation and provides another 
example of the complementarity in searches for physics beyond the 
standard model that a photon collider may provide. 

\subsection{Non-standard gauge boson couplings at photon colliders}

One of the less well-studied properties of the SM are the gauge 
boson self-interactions. Precise measurements would require high 
enough energies to produce at least a pair of $W$ or $Z$ bosons
with large cross sections. At a photon-photon collider such 
processes, and even three- and four-boson production, could be 
studied. It is conceivable that the processes $\gamma \gamma 
\rightarrow W^+ W^-$ or $e \gamma \rightarrow \nu W$ could be 
measured with a precision high enough to be sensitive to two-loop 
contributions. From the theoretical point of view, a full 
understanding of these processes is difficult since it involves 
the yet unsolved problem of a consistent definition of scattering 
amplitudes for unstable particles \cite{ginzburg}. 

The main motivation for a study of multiple boson production 
consists in the possibility to observe deviations from standard 
model predictions. Many speculative models of physics beyond the 
standard model (from more conventional extensions of the underlying 
gauge group to models with extra dimensions) lead to effective 
three- and four-gauge boson vertices with couplings that deviate 
from the standard model prediction. There are various parametrizations 
of these couplings that are conventionally used to perform 
model-independent studies. As a minimal requirement, anomalous 
couplings have to be defined in such a way that the electromagnetic 
$U(1)$ gauge invariance is respected. However, for realistic 
underlying theories that are compatible with existing data, one should 
base studies on an effective Lagrangean that takes also $SU(2)$ 
invariance into account. Limits on anomalous gauge-boson couplings 
have been determined by the LEP experiments and could be considerably 
improved at a linear collider based on measurements of the process 
$e^+ e^- \rightarrow W^+ W^-$. Also the LHC is expected to contribute 
to this kind of analyses since there protons can be used as a source 
of photons and anomalous gauge boson couplings could be studied in 
processes like $pp \rightarrow pp + W^+ W^-$ and $pp \rightarrow pp 
+ ZZ$ (the latter is forbidden at tree-level in the standard model). 
For the LHC, studies taking into account a realistic detector 
environment have obtained limits that improve those obtained by the 
LEP experiments by a factor of up to $10^3$, but these studies assume 
independent variations of coupling parameters not respecting 
$SU(2)$ symmetry \cite{chapon}. A study based on a $SU(2)$-invariant 
effective Lagrangean \cite{manteuffel} has revealed strong 
correlations and smaller sensitivities. Limits that could be obtained 
at a photon collider are comparable with those from an $e^+ e^-$ 
linear collider, but better than those from photon-photon collisions 
at the LHC by a factor of roughly $10^2$.

\section{Direct photons}

The study of direct photon production as a testing ground for QCD 
has a long tradition, also as a topic at this series of conferences. 
A considerable amount of experimental and theoretical work was 
reported already at {\em Photon 2007} \cite{heinrich}. A very good 
description by QCD predictions at NLO has been obtained for inclusive 
prompt photon production in hadronic collisions from the majority of 
experiments (i.e., with the exception of E706) for energies ranging 
from about 20 GeV up to 1.96 TeV and up to transverse momenta slightly 
above 200 GeV. Recent improved measurements of prompt photon + jet 
production by D0 show a slight disagreement of the predicted 
$p_T$-slope. Data for photoproduction from H1 and ZEUS are 
underestimated by present NLO calculations if one considers inclusive 
prompt photon production, even at the largest measured values of the 
photon transverse momentum of $p_T = 10$ GeV \cite{arleo}. In the 
presence of an accompanying jet, however, there appears to be good 
agreement if the photon $p_T$ is large, i.e.\ for $p_T > 7$ GeV. The 
presence of a second hard scale, provided by the jet-$p_T$, seems to 
make predictions of a perturbative calculation more reliable. The 
rapidity distribution is usually not described as good, but this is 
simply so because corresponding data are dominated by small values of  
$p_T$. There are now also data for prompt photon production in the 
deep-inelastic kinematic regime from both experiments at HERA. These 
data, however, can at present be compared with leading-order 
predictions only and it is not surprising that no agreement of data 
with theory can be obtained. 

It is interesting to note that an alternative approach based on the 
so-called quasi-multi-Regge-kinematics \cite{saleev} describes both 
inclusive photon production and the associated production of a photon 
and a jet in deep inelastic scattering at HERA. Moreover, it was 
successfully applied to heavy quark production. In this approach, 
a class of corrections beyond the collinear parton model is resummed 
in the framework of an effective theory and only tree-level 
calculations are needed to obtain predictions. It remains to be 
studied how this approach could be extended to the next-to-leading 
order level, so that a reliable estimation of scale uncertainties 
can be obtained. 

It would be very interesting if additional direct photon data for more 
exclusive measurements could be made available, as for example $\gamma 
+ c/b$-tagged jets, or $\gamma$ plus a rapidity gap. Also in this 
case, more theoretical work is needed. The recent work on photon + jet 
associated production in $pp$ collisions \cite{belghobsi} has shown 
that forthcoming measurements at the LHC are promising and may provide 
sensitivity on parton distribution and fragmentation functions. 

Prompt photon production in nuclear collisions at RHIC is believed 
to be a crucial measurement since it might provide a reference 
process to be compared with jet quenching. Since photons, once 
produced, do not interact strongly with the nuclear medium in 
heavy-ion collisions, a comparison of both processes should help in 
understanding the dynamics of the quark-gluon plasma. The essential 
object needed for reliable predictions of photon production is the 
gluon density in nuclei, but it is poorly constrained by present 
fixed-target data. This lack of precise knowledge leads to huge 
uncertainties at low $x$ and low scales. In addition, effects like
\begin{wrapfigure}[18]{r}{0.65\columnwidth}
\centerline{\includegraphics[width=0.58\columnwidth]%
{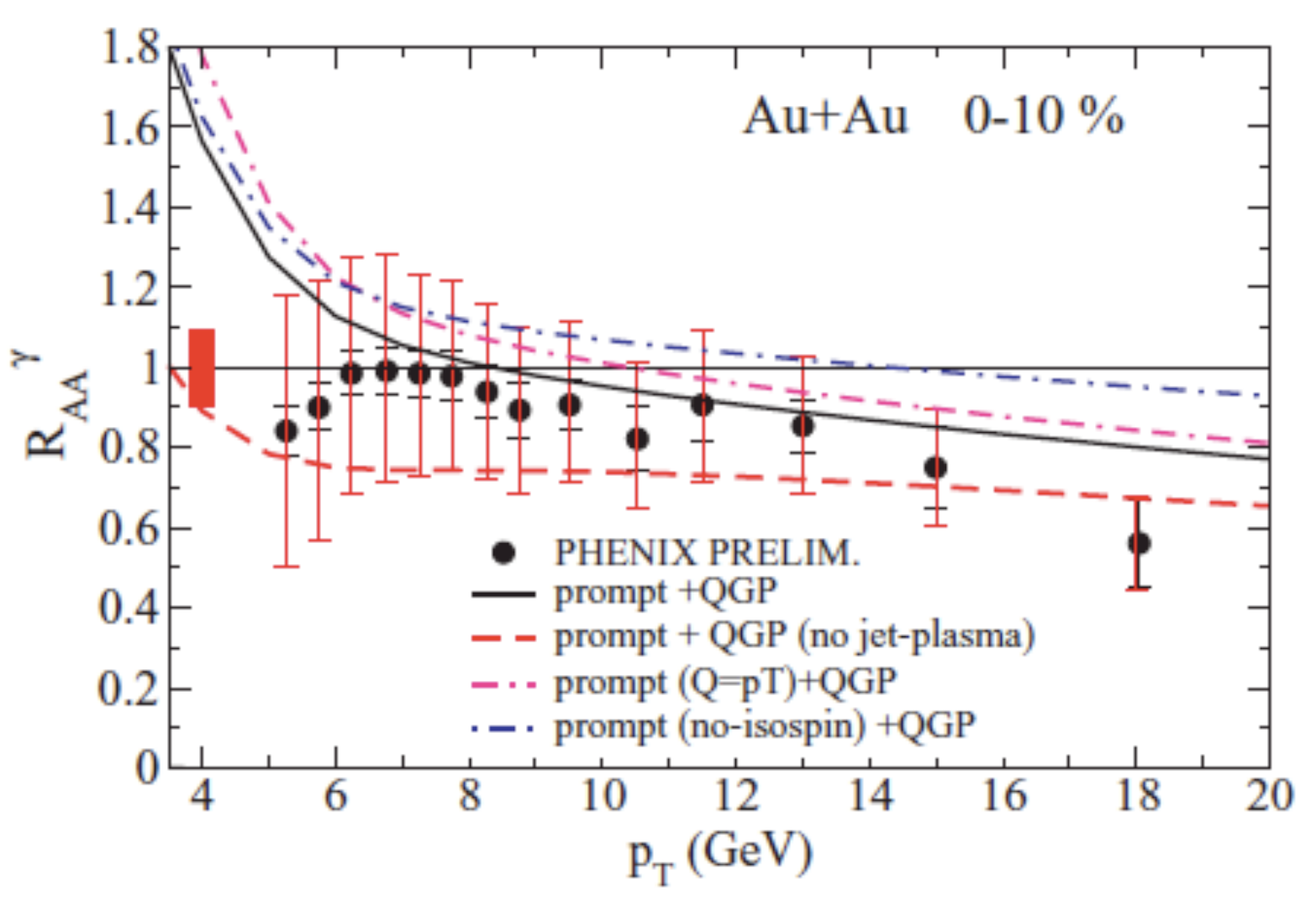}}
\caption{Prompt photon production in $Au+Au$ collisions at PHENIX 
  \cite{tgfh}.}
\label{Fig:hs-3}
\end{wrapfigure}
jet-photon conversion, me\-dium-in\-duced photon emission, or photon 
quenching may be important. Data from PHENIX at RHIC on the nuclear 
production ratio $R_{AA}$ for photons produced in $Au+Au$ collisions 
are consistent with various models predicting almost no suppression 
(see Fig.\ \ref{Fig:hs-3}). Certainly, more precise data are needed; 
but it seems that also the observed problems in the description of 
other prompt photon data, in particular from photoproduction, have to 
be resolved before firm conclusions can be drawn with confidence. 

\section{Photon structure and exclusive hadron production}

Since its first measurement by PLUTO in 1981, the analysis of the 
photon structure function $F_2^{\gamma}$ has been a fruitful 
laboratory of perturbative QCD. The corresponding final LEP2 results 
published in 2005 have provided us with very precise data spanning 
two orders of magnitude in $Q^2$, and more improved data may be 
expected from Belle and BaBar, or possibly from the ILC. Most of the 
data come from deep inelastic scattering off a photon target with 
negligibly small virtuality $P^2$. At large $P^2$, the photon 
structure function $F_2^{\gamma}(x, Q^2, P^2)$ is suppressed; however, 
in the kinematic range $Q^2 \gg P^2 \gg \Lambda^2_{\rm QCD}$ 
perturbative QCD provides a definite prediction, both for the shape 
and the magnitude of $F_2^{\gamma}(x, Q^2, P^2)$ since no separate 
non-perturbative hadronic input is needed as for the case of $P^2 
\simeq 0$. In \cite{sasaki}, heavy quark effects at NLO and target 
mass corrections have been studied for the structure function of a 
highly virtual photon, with the conclusion that PLUTO data at $Q^2 = 
5$ GeV$^2$, $P^2 = 0.35$ GeV$^2$ are better described with 3 massless 
quarks + massive charm than with 4 massless quarks. For L3 data at 
higher $Q^2$ and $P^2$ where $b$ quarks contribute, the comparison 
is not conclusive, both due to the smaller $b$-quark charge and the 
less precise measurements. 

\begin{figure}[h]
\centerline{
\includegraphics[width=0.3\columnwidth]{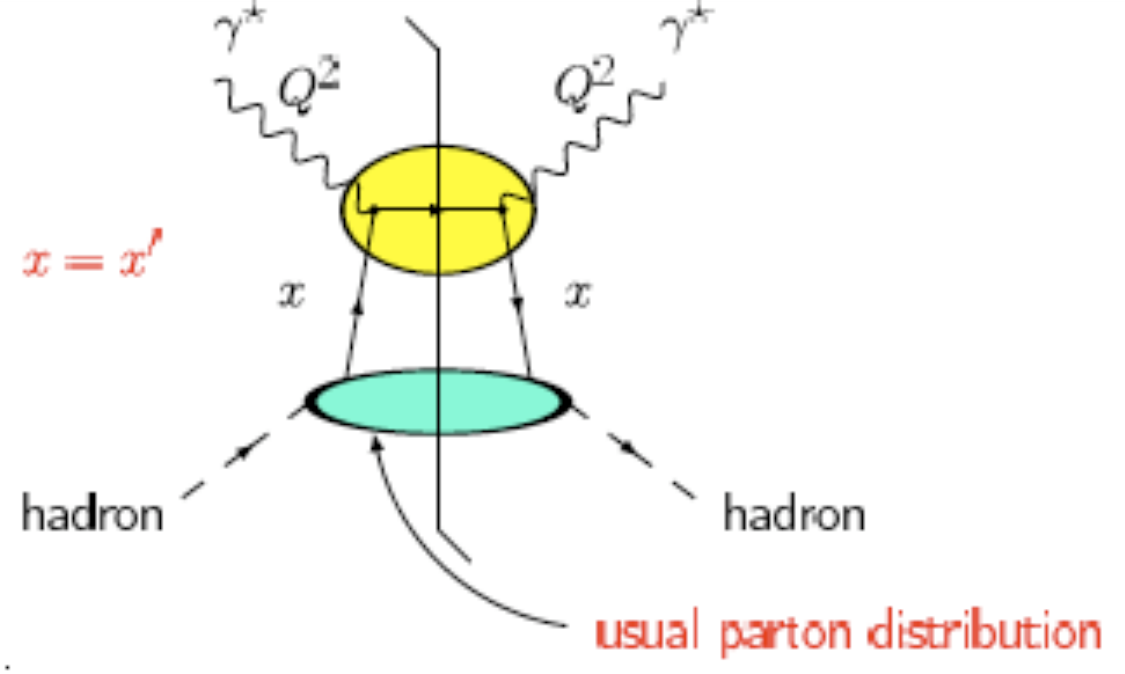}
\includegraphics[width=0.3\columnwidth]{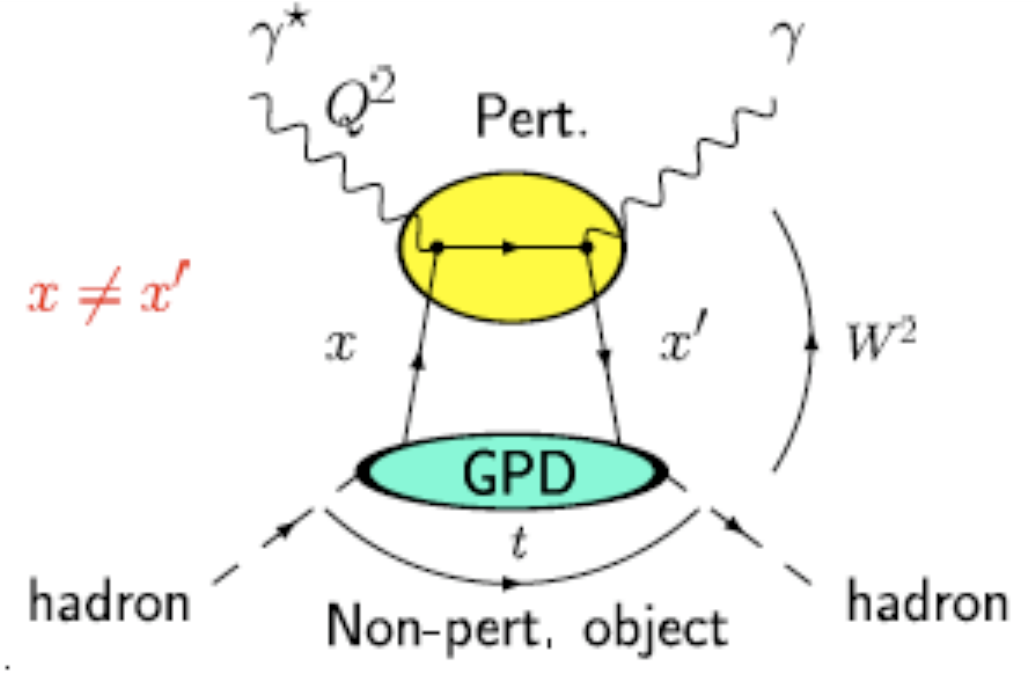}
\includegraphics[width=0.3\columnwidth]{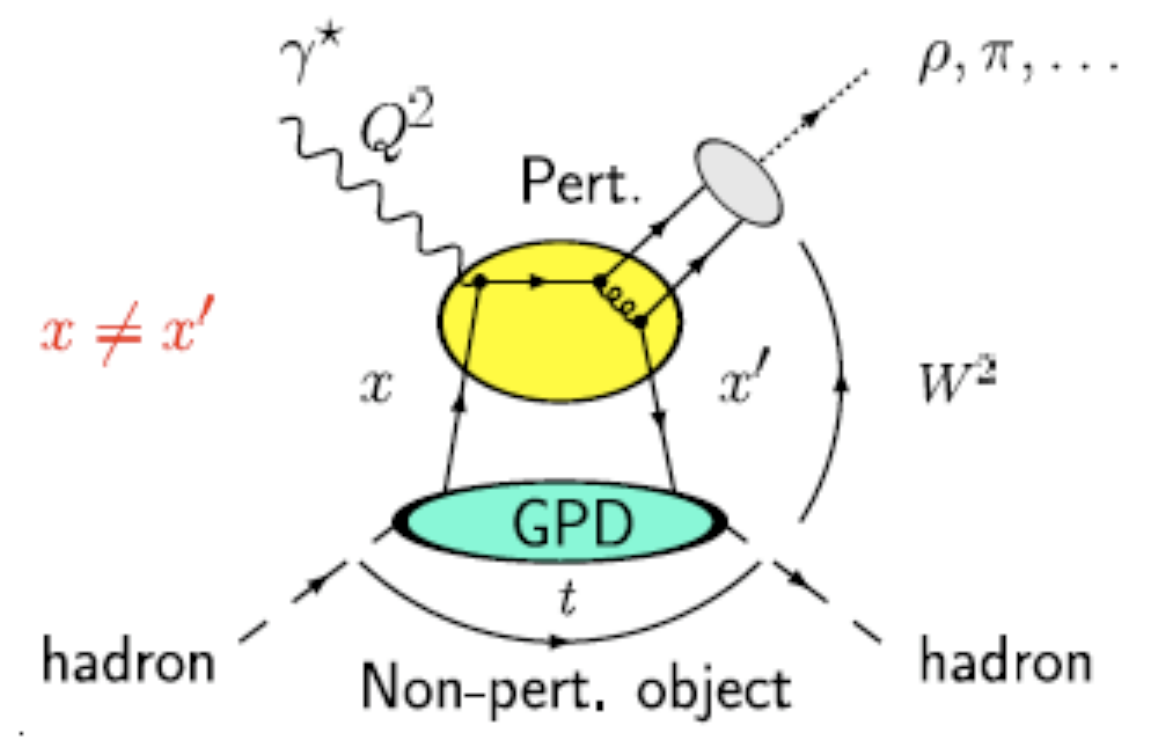}
}
\centerline{
\includegraphics[width=0.3\columnwidth]{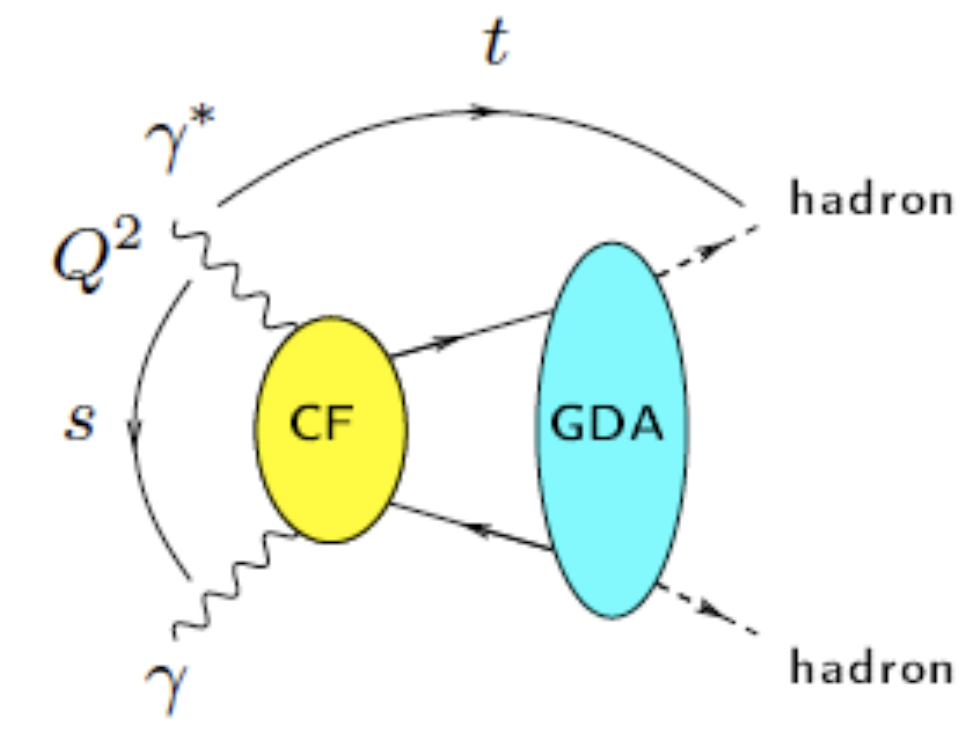}
\includegraphics[width=0.6\columnwidth]{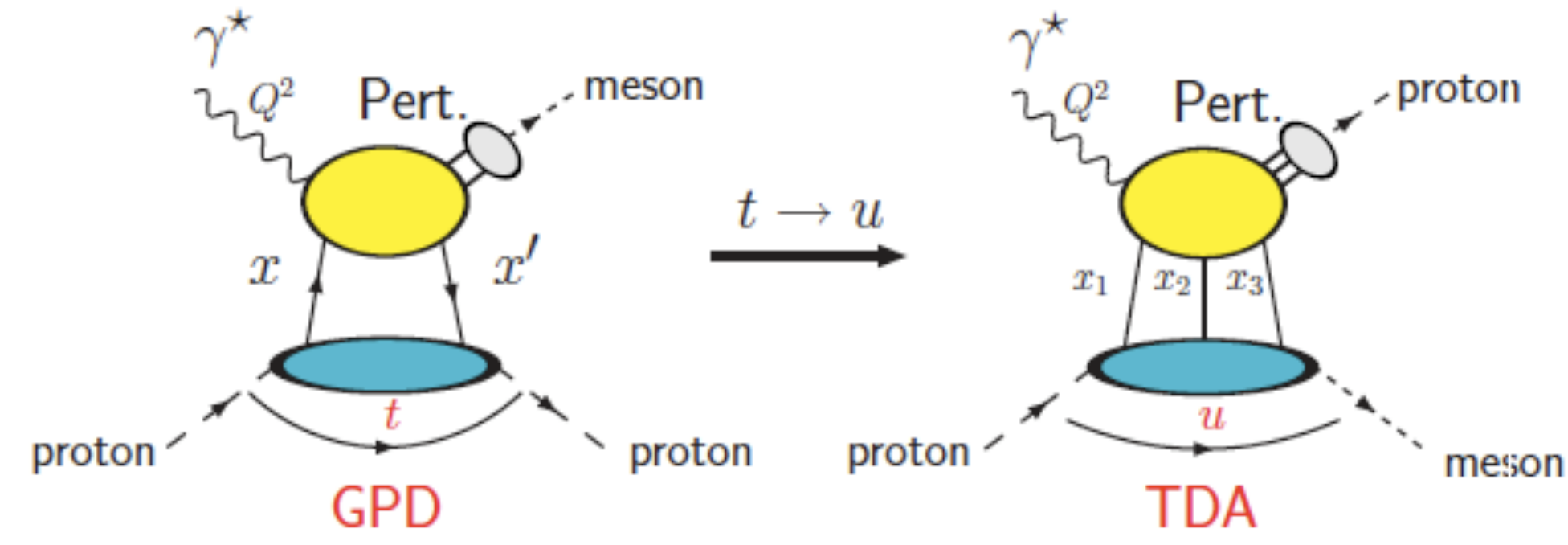}
}
\caption{Basics of hard exclusive processes \cite{wallon}.}
\label{Fig:hs-4}
\end{figure}

The hadronic properties of the electromagnetic current measured in 
deep-inelastic scattering and described with the help of structure 
functions can be studied in the framework of perturbative QCD since 
the factorization theorem allows one to separate structure functions 
into hard parton scattering cross sections and non-perturbative 
hadronic matrix elements of certain operators constructed from 
quark and gluon fields. In the case of inclusive deep-inelastic 
scattering, these matrix elements correspond to the kinematics of 
$\gamma^{\ast}$-hadron forward scattering and are called parton 
distribution functions (PDF). In the case of non-forward scattering, 
as for virtual Compton scattering $\gamma^{\ast} + p \rightarrow 
\gamma + p$, the necessary matrix elements correspond to so-called 
generalized parton distribution (GPD) functions. If the photon in 
the final state of virtual Compton scattering is replaced by a 
hadron, so that exclusive forward-production of hadrons is described, 
one has to introduce in addition distribution amplitudes (DA) or, 
for backward-production after $t \rightarrow u$-channel crossing, 
transition distribution amplitudes (TDA). Finally, $s \rightarrow t$ 
crossing leads from GPDs to generalized distribution amplitudes 
(GDA) needed to describe hadron production in two-photon processes. 
The diagrams shown in Figure \ref{Fig:hs-4} illustrate how the 
various hard exclusive processes and the objects required for their 
description are related to each other. 

Measurements of exclusive processes like deeply virtual Compton 
scattering and meson production, also for experiments with polarized 
beams, are being performed at moderate and high energies and the 
amount and precision of corresponding data require corresponding 
efforts towards an improved theoretical understanding. Considerable 
work in this direction has been done in the recent years and a 
framework has been developed to describe a large number of processes. 
However, factorization has been shown to work only for very few cases 
(e.g.\ $\gamma^{\ast} p \rightarrow \gamma p$), for others a prove at 
any order is not yet available, but factorization is plausible (e.g.\ 
for processes involving TDAs). Some studies have revealed explicit 
factorization breaking (e.g.\ $\gamma^{\ast}_T p \rightarrow \rho_T 
p$ with a transversely polarized photon and $\rho$) \cite{wallon}. 

Corresponding calculations are complicated and the study of simplified 
situations may often be helpful. For example, the process $\gamma 
\gamma \rightarrow \gamma \gamma$ is considered as a theory playground 
to study factorization in GPDs close to threshold kinematics 
\cite{szymanowski}. No phenomenological application for this process 
is visible, but apart from insight into the way how factorization 
works, one may also obtain hints for a reasonable choice of 
parametrizations of GPDs in more realistic cases. Similarly, 
calculations for $\gamma \gamma^{\ast} \rightarrow \pi \pi$ in the 
framework of a euclidean $\phi_E^3$ model have been used to study the 
duality between factorization into GDAs and TDAs \cite{anikin}. In the 
overlap region, when both $s$ and $t$ are small compared to $Q^2$, 
there is an ambiguity since both factorization into GDAs and TDAs can 
operate and it is not clear whether the two descriptions are equivalent 
to each other on a quantitative level, or whether the predictions based 
on both mechanisms should be added. In the scalar $\phi_E^3$ model, 
duality between the two mechanisms has been demonstrated. If this 
property also holds for the case of QCD, then it would pose strong 
restrictions on the allowed non-perturbative ingredients needed for 
the description of various exclusive processes.

The framework for backward production of mesons was described in  
\cite{lansberg}. The kinematics with small $u$ forces one to consider 
matrix elements that describe the exchange of three quarks and 
factorization leads to TDAs, i.e.\ probability amplitudes to find a 
meson inside a nucleon. The non-perturbative TDAs have to be modeled 
by a comparison with measurements, e.g.\ with existing data for 
$\gamma^{\ast} p \rightarrow \pi p$ or $\gamma^{\ast} p \rightarrow 
\eta p$, or based on measurements of related processes like $p\bar{p} 
\rightarrow \gamma^{\ast} \pi^0 \rightarrow \ell^+ \ell^- \pi^0$ or 
$\gamma^{\ast} p \rightarrow J/\Psi p$, which may be accessible by 
future experiments, for example at GSI/FAIR, at JLAB or at 
B-factories.

GDAs are needed to describe hadron-pair production in 2-photon 
processes and can be used also to describe the production of two 
pairs of mesons, e.g.\ in $\gamma \gamma \rightarrow \pi^+ \pi^- 
\pi^+ \pi^-$. This process was identified as a candidate for the 
observation of the perturbative odderon. In the language of 
perturbative QCD, the odderon is described by the exchange of three 
gluons in the color singlet state. A suitably defined angular 
asymmetry for $2\pi$-pair production is sensitive to the interference 
of the odderon and the Pomeron. The $2\pi$ GDAs are unknown and have 
to be modeled; reasonable choices for them predict asymmetries that 
rise above the level of 10\,\% only at very low or very large $2\pi$ 
invariant masses. The event rates in 2-photon scattering at the LHC 
may be large enough, but background from hadronic interactions will 
most likely prevent a corresponding analysis \cite{schwennsen}.

Similarly to conventional parton distribution functions, also DAs 
and their generalizations, the transverse-momentum unintegrated 
light-cone wave functions obey simple evolution equations derived 
in perturbative QCD, at leading order so-called 
Efremov-Radyushkin-Brodsky-Lepage evolution equations. The DAs reach 
an ultraviolet fixed-point at large scales and are, consequently, 
uniquely defined by perturbative QCD; however, scales that are 
accessible by present experiments are low and corresponding 
predictions are sensitive to the non-perturbative initial conditions. 
Therefore, a comparison with data is mainly a test of various model 
assumptions used to obtain these initial conditions, as for example 
models based on QCD sum rules, relativistic quark models, instanton 
liquid models, or effective chiral quark models. The case of pion or 
photon DAs is particularly interesting since results from lattice 
QCD are available, while experimental information is obtained only 
indirectly, extracted from hard di-jet production by incident pions 
or real photons. In the case of pions, the DAs (and corresponding 
generalized form factors) obtained from a chiral quark model and 
evolved at leading order \cite{broniowski} are in reasonable overall 
agreement with the available data and results from lattice simulations. 

The classical area for the study of GPDs is deeply virtual Compton 
scattering (DVCS), i.e.\ the process $e^{\pm} N \rightarrow e^{\pm} 
N \gamma$. Other related processes are the production of lepton 
pairs, e.g.\ $e^{\pm} N \rightarrow e^{\pm} N \mu^+ \mu^-$ or 
$\gamma N \rightarrow N^{\prime} e^+ e^-$ and hard exclusive meson 
production, e.g.\ $e p \rightarrow ep \pi^0$, $e p \rightarrow ep 
\rho$, $e p \rightarrow en \pi^+$, $e p \rightarrow en \rho^+$. 
There is a considerable amount of data available now, for example 
from H1, ZEUS and HERMES at HERA, or from the CEBAF Large Acceptance 
Spectrometer (CLAS) and Hall A (CEBAF Experiment 91-010) 
collaborations at JLAB. Moreover, measurements are expected to 
provide more data from COMPASS, or from experiments at JLAB with a 
future 12 GeV beam upgrade. 

Before fitting GPDs to data, their functional form has to be carefully 
modeled. GPDs are intricate functions with a non-trivial interplay of 
the dependence of various kinematic variables and subject to 
constraints in a number of limiting cases. For example, they have to 
reduce to the conventional PDFs in the limit of forward scattering, 
there are various sum rules, and at LO they have to obey a positivity 
constraint. Moreover, their evolution at next-to-leading order has to 
be described properly. Because of their relation with conventional 
PDFs, it seems reasonable to perform a simultaneous fit of inclusive 
DIS and DVCS data. If based on a flexible ansatz for the GPDs, good 
fits to DVCS data can be obtained \cite{dmueller}. The complicated 
structure of GPDs make them a promising tool to reveal the transverse 
distribution of partons in a nucleon, or to address the spin content 
of the nucleon. It will be interesting to perform at more detail 
comparisons with non-perturbative methods, as for example lattice 
simulations. 

\section{Small-$x$ and total cross sections}

At small Bjorken-$x$, the behavior of structure functions in 
deep inelastic scattering is most conveniently studied in terms 
of color dipoles, the pair of color charges carried by 
quark-anti-quark pairs into which the virtual photon splits during 
its interaction with a nucleon. At high energies, these color charges 
propagate along straight lines separated by a transverse distance 
that remains unchanged, and the associated color dipole is described by 
a two-Wilson-line operator. The low-$x$ evolution of the structure 
functions is then governed by the rapidity evolution of the color 
dipoles. At leading order in the leading logarithmic approximation, 
the rapidity evolution can be determined by the non-linear 
Balitsky-Kovchegov (BK) equation. The BK equation extends the BFKL 
equation by a non-linear term which describes parton annihilation and 
predicts saturation at low $x$. The leading-order BK equation is 
conformally invariant, however, at next-to-leading order, conformal 
invariance is broken in QCD by the running of the strong coupling 
constant. 

It should be instructive to consider the evolution of color dipoles 
in a ${\cal N} = 4$ supersymmetric Yang-Mills theory. The ${\cal N} 
= 4$ SYM theory has received a lot of interest recently as it is 
conjectured to be dual, at strong coupling, to a type IIB string 
theory in 5-dimensional Anti-de-Sitter space-time. Its $\beta$-function 
vanishes and one can hope that its conformal invariance provides 
strong restrictions on the correct effective action of QCD at high 
energies. At LO, the color dipole evolution equation has the same 
form as in QCD and it has been demonstrated \cite{chirilli} that with 
suitably defined "composite conformal dipole operators", the 
conformally invariant analytic part of the QCD evolution equation at 
NLO can be obtained in that theory. It has also been shown that the 
resulting evolution equation agrees with the forward BFKL equation at 
next-to-leading order. 

The complicated physics of the deep inelastic structure functions 
results from the cooperation of weak coupling due to the asymptotic 
freedom of QCD at high energies and the high density of gluons 
inside the nucleus. The gluon density increases towards lower 
Bjorken-$x$ since at weak coupling, soft and collinear gluon emission 
is favored. Ultimately, this would break unitarity. However, gluon 
bremsstrahlung is limited by recombination effects, an effect which 
restores unitarity and leads to saturation at a limit depending on 
$x$ and $Q^2$ given by $Q_s^2(x) \propto 1 / x^{\lambda}$ with 
$\lambda \simeq 0.2 - 0.3$. 

The situation may change dramatically in heavy-ion collisions where 
data indicate the presence of strong coupling effects. According to 
the famous conjecture by Maldacena, there is a correspondence between 
a strongly coupled gauge theory and a string theory at weak coupling. 
If this conjecture is right, it would be possible to infer from the 
study of the classical dynamics of a black hole in AdS$_5$-supergravity 
properties of photon interactions with a strongly coupled quark-gluon 
plasma. Corresponding investigations \cite{iancu} lead to the 
conclusion that saturation should set in faster and the behavior of 
$Q_s^2(x)$ would be changed to $Q_s^2(x) \propto 1 / x$. As an 
important consequence, since strong coupling would lead to unlimited, 
quasi-democratic parton branchings, one would expect to observe no 
collimated jets in $e^+e^-$ annihilation, and no large-$x$ partons in 
the hadron wave functions, i.e.\ no jets in the forward or backward 
direction of hadron-hadron collisions. It has to be seen whether 
future data from RHIC can contribute to a clarification of these 
concepts. 

\section{Vacuum polarization and $g-2$}

The comparison of experimental data with theoretical predictions 
for the anomalous magnetic moment of the muon has received 
considerable interest in the past years, after the experiment 
E821 at Brookhaven has published the most precise measurement
of $a_{\mu}^{\rm exp} = (g_{\mu}-2)/2 = (11 659 208.0 \pm 6.3) 
\times 10^{-10}$. Since then a discrepancy of about $3\sigma$ has 
been observed. It is an open question whether this is a sign of new 
physics, or of an incomplete understanding of photon-hadron 
interactions contributing to $a_{\mu}$. 

\begin{wrapfigure}{r}{0.53\columnwidth}
\centerline{\includegraphics[width=0.45\columnwidth]
{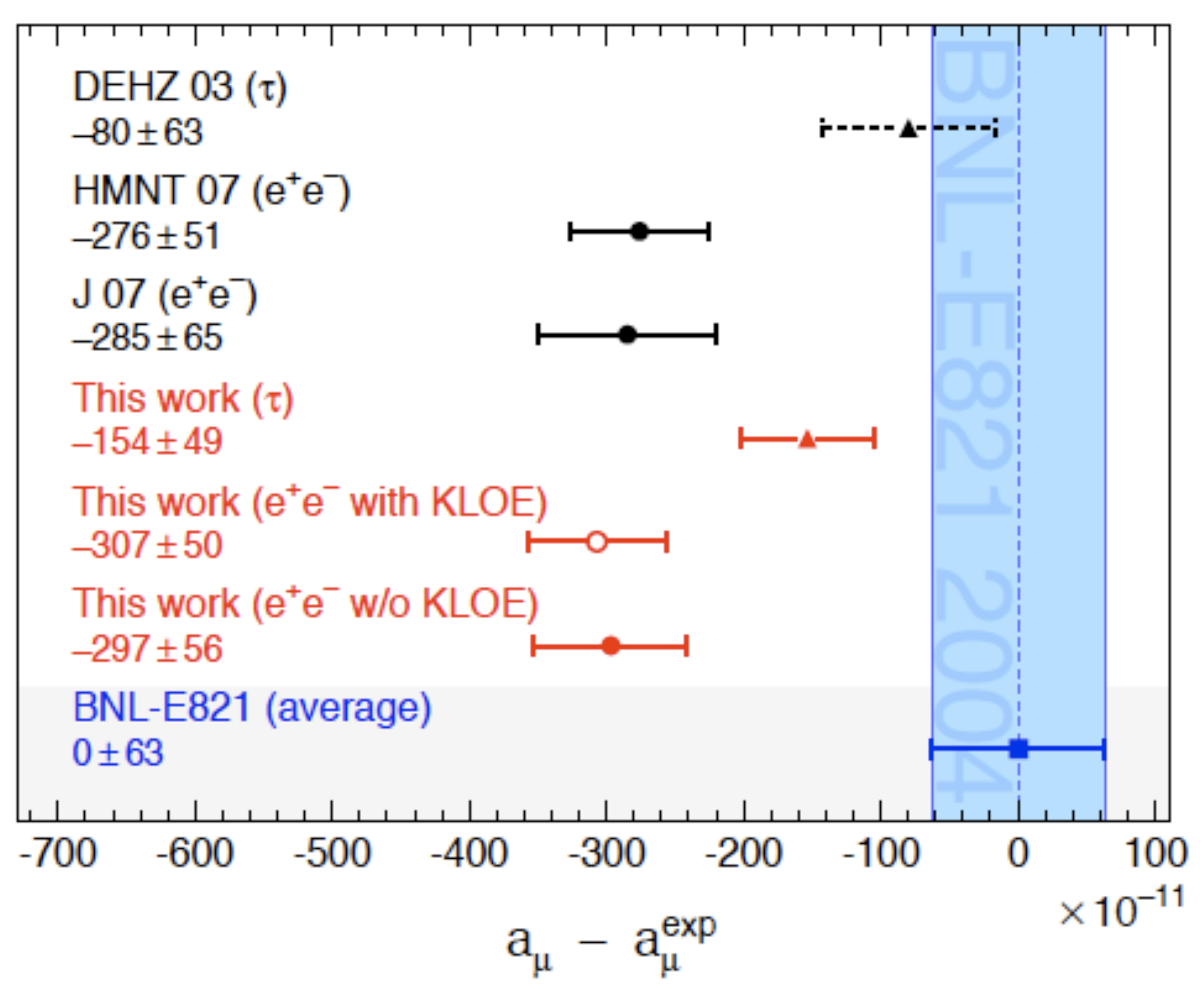}}
\caption{Status of $g-2$ at {\em Photon 2009} \cite{D-0906}.}
\label{Fig:hs-5}
\end{wrapfigure}

At present, the precision of theoretical calculations nicely match 
the size of the experimental uncertainty, but for a conclusive 
comparison of theory with results of future experiments aiming at 
an accuracy of $\pm 1.5 \times 10^{-10}$, considerable improvements 
in the theoretical understanding will be necessary. The present 
theoretical uncertainty is dominated by the precision with which 
hadronic contributions to $a_{\mu}$ can be calculated. 

Hadronic contributions enter at the two-loop level via the hadronic
vacuum polarization (the two-point correlator of the electromagnetic
current) and at the three-loop level via light-by-light scattering (a
certain component of the four-point correlator of the electromagnetic
current). The former contribution, denoted $a_{\mu}^{\rm had,LO}$, was
traditionally calculated, by using a dispersion relation, from data of
the total hadronic cross section for $e^+ e^-$ annihilation and its
accuracy is limited by those data.  Improvements in the measurement of
$e^+ e^-$ cross section data by the CMD-2, SND and KLOE collaborations
have reduced the accuracy of $a_{\mu}^{\rm had,LO}$ to $\pm 5.0 \times
10^{-10}$ \cite{D-0906}.  On the other hand, with the increasing
precision of LEP data for the decay spectrum of the $\tau$, it became
possible to use, instead of $e^+ e^-$ data, the $\tau$ spectral
functions in the calculation of $a_{\mu}^{\rm had,LO}$. The recent
analysis \cite{D-0906} reported at this conference finds a precision of
$\pm 4.9 \times 10^{-10}$, very similar to the $e^+ e^-$-based
calculations. It takes into account recent high-precision data from the
Belle collaboration on the decay $\tau^- \rightarrow \pi^- \pi^0
\nu_{\tau}$ and a new analysis of isospin-violating corrections.
Interestingly, this analysis results in a better agreement with $e^+
e^-$-based calculations than previously, but is shifted away from
$a_{\mu}^{\rm exp}$ by $-15.4 \times 10^{-10}$ (1.9$\sigma$). The
long-standing discrepancy between spectral functions for $e^+ e^-
\rightarrow \pi^+ \pi^-$ and $\tau^- \rightarrow \pi^- \pi^0 \nu_{\tau}$
is reduced, but still present, in particular for data from KLOE. Very
recent\footnote{This was reported at the XXIV International Symposium on
  Lepton and Photon Interactions at High Energies, Hamburg, 17-22
  August, 2009, by B.\ Shwartz.}  data from BaBar \cite{babar09} on $e^+
e^- \rightarrow \pi^+ \pi^- (\gamma)$ using the ISR-method, however,
have been used to calculate $a_{\mu}^{\rm had,LO}$ \cite{dhmyz}, leading
to a shift of $10.6 \times 10^{-10}$ towards the experimental value.
This brings this $e^+e^-$-based analysis in rough agreement with the
$\tau$-based value (at 1.4$\sigma$).  The discrepancy between the most
recent analysis including reevaluated $\pi^+\pi^-$ and
$\pi^+\pi^-2\pi^0$ data \cite{dhmyz} and the $g-2$ measurement is
3.1$\sigma$. More precision data from BaBar for the $4\pi$-channel are
expected soon and will hopefully contribute to a further clarification
of the situation.

\begin{wrapfigure}{r}{0.43\columnwidth}
\centerline{
\includegraphics[width=0.2\columnwidth]%
{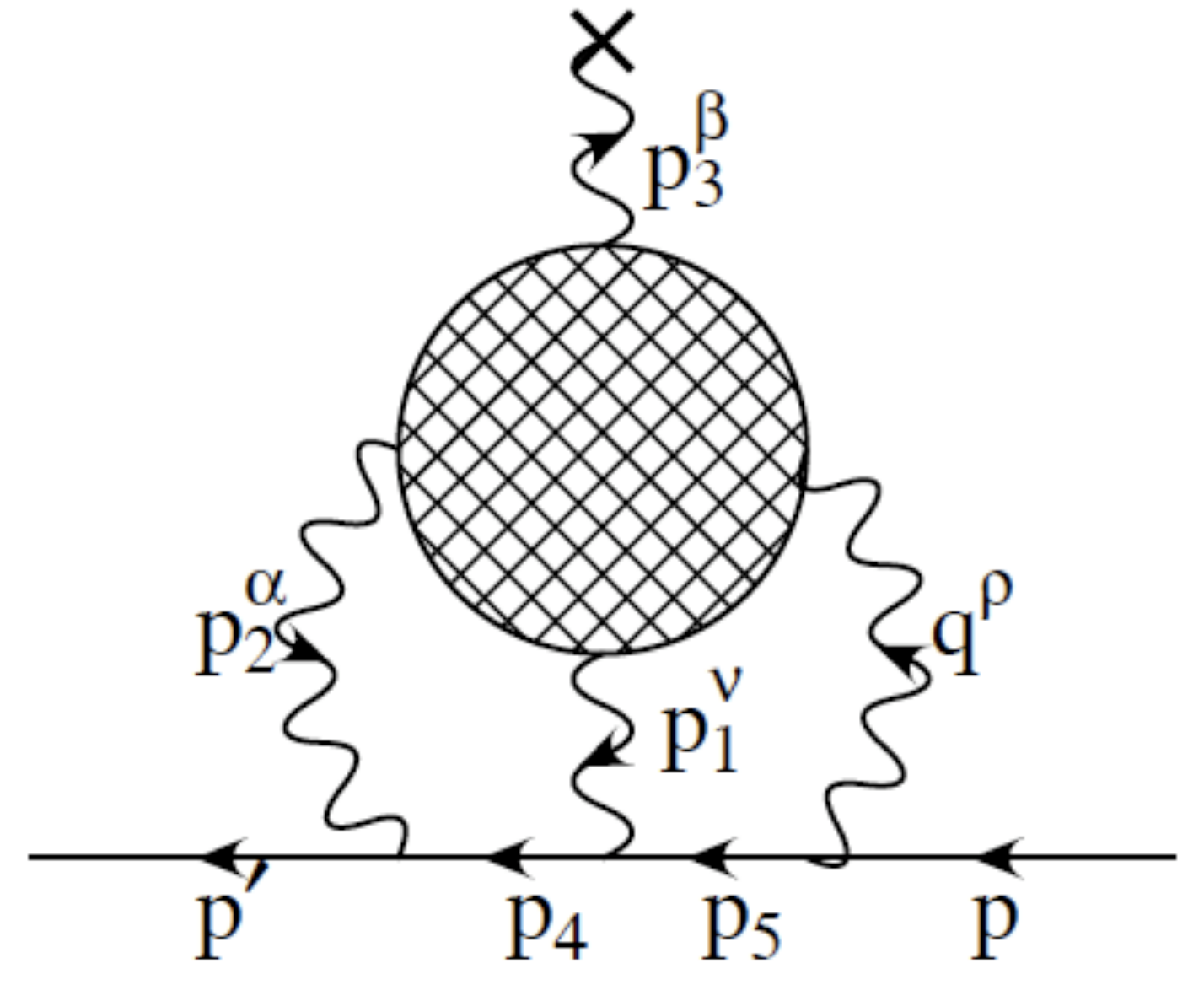}
\raisebox{1.9mm}{\includegraphics[width=0.2\columnwidth]%
{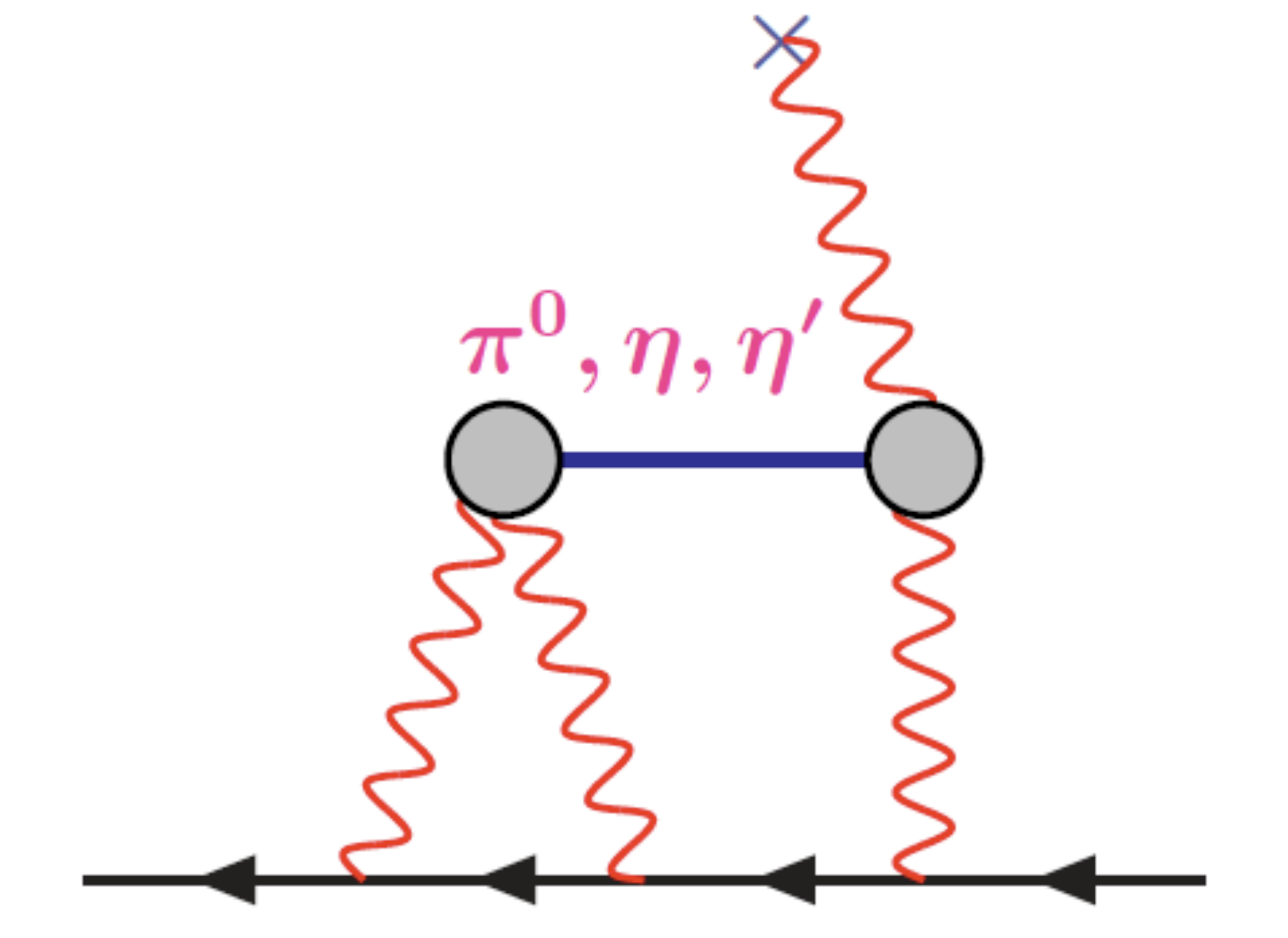}}
}
\caption{Hadronic light-by-light scattering and the dominating 
  $\pi^0, \eta, \eta^{\prime}$ exchange diagram \cite{prades}.}
\label{Fig:hs-6}
\end{wrapfigure}

The hadronic contribution to $a_{\mu}$ from light-by-light 
scattering, $a_{\mu}^{\rm hLbL}$, cannot be directly related to 
experimental data and there is at present no complete calculation 
from first principles. A counting of various contributions can be 
based on the $1/N_c$ expansion and power-counting of chiral 
perturbation theory. At order $O(N_c)$, one-particle-reducible 
exchange diagrams involving Goldstone bosons (at $O(p^6)$) and 
non-Goldstone bosons (at $O(p^8)$) dominate, whereas loop-diagrams 
start at order $O(1/N_c)$. Various model calculations are in fair 
agreement with each other, resulting in the dominating contribution 
$a_{\mu}^{\rm hLbL}(\pi^0, \eta, \eta^{\prime}) = (11.4 \pm 1.3) 
\times 10^{-10}$ \cite{prades} from Goldstone-boson exchange. 
Contributions due to the exchange of pseudo-vector and scalar 
resonances, heavy quark loops and $\pi$ loops are less well-known 
and cancel each other to some extent. A conservative estimate of 
their precision, but adding theoretical errors from different 
contributions in quadrature, results in $a_{\mu}^{\rm hLbL} = (10.5 
\pm 2.6) \times 10^{-10}$ \cite{prades}. Improvements will be 
difficult to achieve, but may be expected by a more refined 
consideration of short-distance QCD constraints and by utilizing 
experimental information on the properties of the off-shell 
$\pi\gamma\gamma$ and $\pi\pi\gamma\gamma$ form factors. Data from 
BaBar, KLOE-2 and DA$\Phi$NE on radiative decays of pions and 
pseudo-vector resonances and other 2-photon processes will play an 
important role in this respect.  

\section{Acknowledgments}
I thank the speakers of the conference for help in preparing this 
summary talk and the organizers for their kind invitation.


\begin{footnotesize}

\end{footnotesize}

\end{document}